\documentclass[
10pt,twocolumn,prl,aps,superscriptaddress,
floatfix,nobalancelastpage,preprintnumbers,reprint,
footinbib,noshowpacs,longbibliography
]{revtex4-2}
\usepackage[dvipsnames]{xcolor}
\usepackage[colorlinks=true,breaklinks=true]{hyperref}
\hypersetup{allcolors=[rgb]{0.0 0.0 0.70},linkcolor=blue}

\usepackage{orcidlink}
\usepackage{microtype}

\usepackage{graphicx}
\usepackage{dcolumn}
\usepackage{esvect}
\usepackage{amsmath}
\usepackage{amsfonts,amssymb}
\usepackage{mathtools}
\usepackage{multirow}
\usepackage{bm}
\usepackage[utf8]{inputenc}
\usepackage{makecell}

\makeatletter

\newcommand{\checknextarg}{\@ifnextchar\bgroup{\gobblenextarg}{}}
\newcommand{\gobblenextarg}[1]{\,\mathrm{#1}\@ifnextchar\bgroup{\gobblenextarg}{}}
\makeatother

\begin{document}

\title{Neutrinos Probe Exotic Energy Losses from the Sun}

\author{Pablo Mart{\'i}nez-Mirav{\'e}~\orcidlink{0000-0001-8649-0546}}
\email{pablo.mirave@nbi.ku.dk}
\affiliation{%
Niels Bohr International Academy and DARK, Niels Bohr Institute, University of Copenhagen,
Blegdamsvej 17, 2100 Copenhagen, Denmark
}%
\author{Irene Tamborra~\orcidlink{0000-0001-7449-104X}}
\email{tamborra@nbi.ku.dk}
\affiliation{%
Niels Bohr International Academy and DARK, Niels Bohr Institute, University of Copenhagen,
Blegdamsvej 17, 2100 Copenhagen, Denmark
}%

\date{\today}

\begin{abstract}

Solar neutrinos provide a direct window into the Sun core, serving as powerful probes of new physics. We construct evolutionary models of the Sun incorporating general, temperature-dependent exotic energy losses. Profiling over solar uncertainties, we confront these models with the latest \textit{pp}, $^7$Be, $^8$B, and CNO neutrino flux measurements alongside helioseismic data. At $3\sigma$, we constrain exotic energy-loss normalization to $\lesssim 0.25$~erg~g$^{-1}$~s$^{-1}$ for a  $(T/10^7 \rm K)^\beta$ scaling, with $0\leq \beta \leq 10$. Combining precision stellar modeling with neutrino observations establishes an effective framework to test beyond-Standard-Model cooling mechanisms, including hidden photons and axions.

\end{abstract}

\maketitle

\textit{Introduction.---}Stars are unique laboratories for fundamental physics~\cite{Raffelt:1996wa,Raffelt:1999tx}. 
Extensions of the Standard Model predict the existence of new particles responsible for enhanced energy loss from the core of stars, therefore modifying stellar evolution and leaving observable imprints on the multi-messenger emission~\cite{Raffelt:1996wa}. 

Solar neutrinos can   probe  non-standard energy losses~\cite{Gondolo:2008dd,Redondo:2013lna,Vinyoles:2015aba}. Produced in thermonuclear reactions, the neutrino flux is sensitive to the temperature and composition of the solar core. The $pp$ flux scales mildly with temperature ($\sim T^{-1}$), whereas the $^7$Be and $^8$B fluxes exhibit much stronger dependences  ($\propto T^{10-11}$ and $T^{24-25}$, respectively)~\cite{Bahcall:1996vj}. The CNO neutrino flux steeply scales  with temperature ($\sim T^{20-23}$) and is  sensitive to the abundances of carbon and nitrogen~\cite{Bahcall:1996vj,BOREXINO:2022abl,BOREXINO:2023ygs}. Over the past decades, solar neutrino experiments have improved the precision of their measurements. The \textit{pp}, $^7$Be, and $^8$B fluxes  provide increasingly stringent tests of solar models~\cite{Borexino:2017rsf,Borexino:2017uhp}, while the  measurement of CNO neutrinos has shed light on the contribution of the CNO cycle to solar energy production~\cite{BOREXINO:2023ygs}. 

Helioseismology  probes the solar interior by measuring the oscillation frequencies of acoustic waves (pressure waves or p-modes) and other resonant vibrations (e.g., g-modes and f-modes), which allow us to reconstruct the radial sound-speed profile~\cite{Rhodes:1997}. Additional seismic diagnostics constrain the radius of the base of the convective envelope~\cite{Basu:1997} and the surface helium abundance~\cite{Basu:2004zg}.
Exotic energy losses can modify the solar temperature and density profiles. Hence,  seismic observables are expected to track exotic energy losses~\cite{Schlattl:1998fz}.
However, the  ``solar modeling problem'' plagues our understanding of the Sun,  i.e.~the existence of persistent discrepancies between solar models and helioseismic observations, most notably in the sound-speed profile, the radius of the base of the convective envelope, and the surface helium abundance~\cite{Amarsi:2026,Buldgen_2025, Buldgen:2024}. Uncertainties in the solar chemical composition and radiative opacities hinder our understanding of the solar interior~\cite{Christensen-Dalsgaard:2008idi,Boothroyd:2002qw,Basu:2007fp, Chen:2024ouz}. This puzzling problem raises the question: can   exotic energy-loss mechanisms  modify the solar structure, improving the agreement between solar models and observations? 

\begin{figure}[t]
    \centering
    \includegraphics[width=\linewidth]{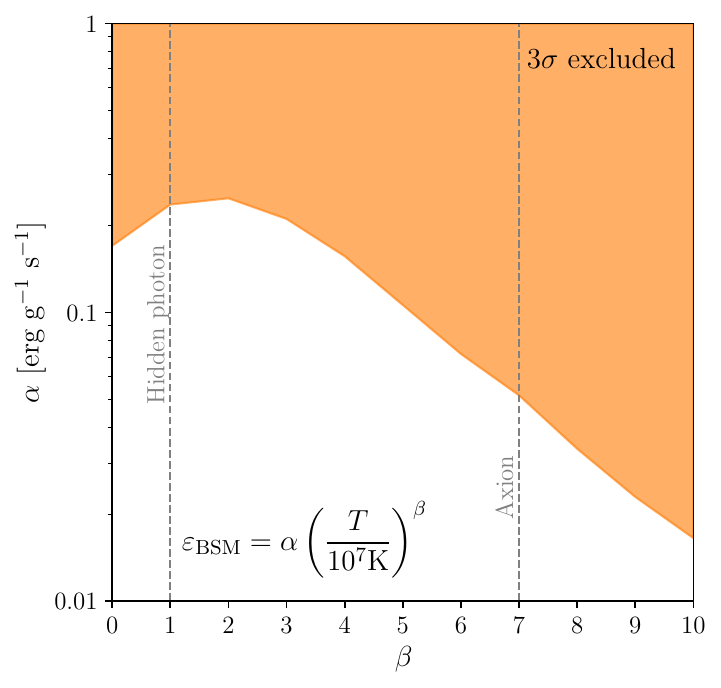}
    \caption{Exclusion bounds at $3\sigma$ (in orange) on exotic energy losses from the Sun, assuming $\epsilon_{\rm{BSM}} \propto \alpha T^\beta$. We evolve advanced solar models and contrast their outcome with solar neutrino flux  measurements ($pp$, $^7$Be, $^8$B, and CNO neutrinos)~\cite{Gonzalez-Garcia:2023kva,BOREXINO:2023ygs},  helioseismic data~\cite{Basu:1997,Basu:2004zg,Rhodes:1997}, surface photometry~\cite{Grevesse:1998bj} as well as measurements of the solar luminosity and its effective temperature~\cite{IAU:2015fjh}. These constraints also apply to   axions ($\beta = 7$)~\cite{CAST:2017uph} and  hidden photons ($\beta = 1$)~\cite{Schwarz:2015lqa}.}
    \label{fig:exclusion}
\end{figure}

In this {\it Letter}, we present a novel end-to-end framework to assess the existence of exotic energy losses in the Sun of the form $\epsilon \propto \alpha T^{\beta}$. We obtain an upper bound on exotic energy loss of $\lesssim 0.25$~erg~g$^{-1}$ (or luminosity emitted in  exotic particles, in units of the solar photon luminosity:  $L_{\rm BSM} \lesssim 0.10~L_\odot$) for temperature dependence ranging from constant up to $T^{10}$, as summarized in Fig.~\ref{fig:exclusion}.
We do so by evolving solar models  with the software \texttt{Modules for Experiments in Stellar Astrophysics (MESA)}~\cite{Paxton:2011,Paxton:2013pj, Paxton:2015jva,Paxton:2019,  MESA:2022zpy}, accounting for  changes in stellar structure and nuclear burning throughout stellar evolution due to exotic energy losses. 
We assess the  resulting  predictions of solar observables from such models against the latest solar data.

Combining measurements of the different components of the solar neutrino flux reveals to be  particularly valuable, as each component probes different physical conditions in the solar interior. 
Moreover, the use of neutrino and helioseismological probes reveals to be crucial in disentangling changes in the solar thermal structure from uncertainties in its chemical composition.
Ours is the first systematic end-to-end treatment of stellar-model uncertainties in constraining exotic energy losses using  neutrino observations.  We propagate uncertainties in the initial stellar parameters and solar age throughout stellar evolution, accurately capturing the non-linear feedback of exotic energy loss on solar structure, nuclear burning, and current observables.

\textit{Solar evolution with non-standard energy losses.}---We compute $\gtrsim 13,000$ spherically symmetric models of main-sequence stars with mass of $1\, M_\odot$ using the stellar evolution code \texttt{MESA}~\cite{Paxton:2011,Paxton:2013pj, Paxton:2015jva,Paxton:2019,  MESA:2022zpy}, v.28.08.1. We adopt a mixing-length theory treatment for convection~\cite{Paxton:2011,Henyey:1965} and account for convective overshooting as in Ref.~\cite{Herwig:2000sq}. We employ the nuclear reaction network \texttt{mesa\_45}, which includes all relevant isotopes and reactions for main sequence stars.

Stellar evolution is governed by energy conservation:
\begin{equation}
\frac{{\rm d}L}{{\rm d}m}
=
\varepsilon_{\rm nuc}
- \varepsilon_{\nu}
- \varepsilon_{\rm grav}
- \varepsilon_{\rm BSM}\, ,
\label{eq:energy-conservation}
\end{equation}
where $L$ is the luminosity and $m$ is the enclosed mass coordinate. The source and sink terms  of energy per unit mass on the right-hand side are  the specific nuclear energy generation rate ($\varepsilon_{\rm nuc}$),  the  energy-loss rate due to neutrinos ($\varepsilon_{\nu}$), and the gravothermal energy term ($\varepsilon_{\rm grav}$, accounting for changes in the internal and gravitational energy associated with the contraction or expansion of the star). The term $\varepsilon_{\rm BSM}$  describes energy losses through exotic processes and is parameterized as follows:
\begin{equation} 
    \varepsilon_{\rm BSM} = \alpha \left(\frac{T}{10^7 {\rm K}}\right)^\beta\, ;
    \label{eq:parametrisation}
\end{equation} 
here $\alpha$ (in units of $\mathrm{erg\,g^{-1} \,s^{-1}}$) indicates the energy loss rate per unit mass, and $\beta$ is responsible for the scaling with temperature. We consider a reference temperature of $10^7$~K as it is close to (yet slightly lower than) the temperature of the solar core. We implement $\varepsilon_{\rm BSM}$  using the \texttt{MESA} hook \texttt{other\_energy\_implicit}. 

We consider a grid with $11$ linearly spaced values of $\beta$ in the range $[0, 10]$ and $13$ logarithmically spaced values of $\alpha$ in the range $[10^{-3}, 1]$, in addition to the case with \mbox{$\alpha =0$} which represents a scenario without exotic energy losses. For each $(\beta,\alpha)$ in this grid (see also  Fig.~\ref{fig:exclusion}), we evolve a set of  solar models. 

\begin{figure}
    \centering
    \includegraphics[width=\linewidth]{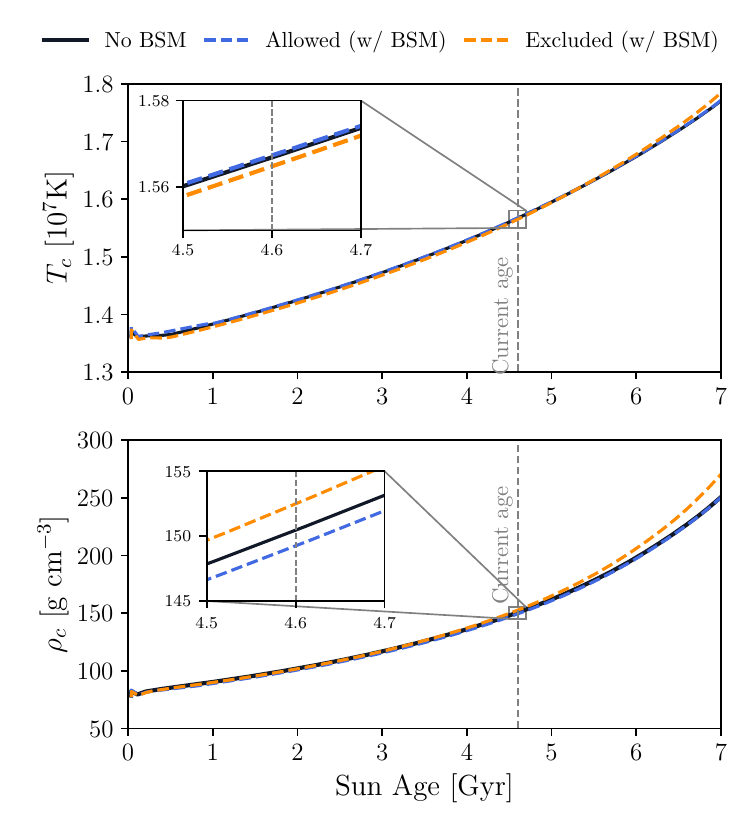}
    \caption{Evolution of the central temperature and baryon density (top and bottom, respectively) for  selected combinations of the parameters $\alpha$ and  $\beta$ in the parameter space displayed in Fig.~\ref{fig:exclusion}: model  allowed  (blue, $\beta = 3$ and $\alpha = 0.1$~erg~g$^{-1}$~s$^{-1}$) and excluded   (orange, $\beta = 7$ and $\alpha = 0.1$~erg~g$^{-1}$~s$^{-1}$) from solar data, as well as  standard solar evolution (black, $\alpha=0$). The insets highlight  differences across models at the the current age of the Sun. 
    }
    \label{fig:evolution}
\end{figure}
The  energy sink term $\epsilon_{\rm{BSM}}$ modifies the thermal evolution of the Sun and its structure, as shown in Fig.~\ref{fig:evolution}. 
The top panel displays the evolution of the central temperature ($T_c$, defined as the temperature at the innermost mass cell in \texttt{MESA}) as a function of the solar age. The central temperature provides a useful proxy for the thermal state of the solar core: an increase in $T_c$ is accompanied by an increase in the temperature throughout core, where solar neutrinos originate. 
Only very small changes in the temperature of the stellar core can mimic the  solar properties observed today in the presence of $\epsilon_{\rm{BSM}}$. 
The baryon density profile (bottom panel) also changes non monotonically.
Such changes imply adjustments in the   nuclear burning rates. Due to the marked dependence of the neutrino flux on $T_c$, small modifications in the temperature profile largely affect the neutrino production rates, as detailed in the  Supplemental Material. Therefore, according to the values of $\beta$ and $\alpha$ in Eq.~\ref{eq:parametrisation}, some solar models fail to reproduce  solar observables (cf.~orange lines in Fig.~\ref{fig:evolution}). Interestingly,  it is also possible to construct solar models in  agreement with  observations but that exhibit different past and future evolution.

The  model predictions depend on the initial composition of the Sun, which we parameterize in terms of the initial helium mass fraction ($Y$), the  carbon, nitrogen, and oxygen fraction ($Z_{\rm CNO}$), and the one of other heavier metals ($Z_{\rm other}$). We set the relative ratio of the isotopes in each of the two groups according to the high-metallicity standard solar model of Grevesse and Sauval  (GS98)~\cite{Grevesse:1998bj} (note that Ref.~\cite{Buldgen:2024aak} recently employed a new abundance set and up-to-date atomic data, reproducing the high-metallicity features of GS98 while addressing the sound-speed discrepancy; yet, these models remain in tension with low-metallicity alternatives~\cite{Asplund:2009fu,2021A&A...653A.141A}). The fraction of initial hydrogen is given by  $X+Y+Z_{\rm CNO}+Z_{\rm other}=1$. Our models  also depend on the  mixing length parameter ($\alpha_{\rm MLT}$) and the overshooting parameter ($f_{\rm ov}$), which account for convective effects in spherically symmetric simulations.

\textit{Statistical analysis.}---To constrain additional forms of energy loss in the Sun, we compare the latest  solar data  with the predictions from our solar models. 
Table~\ref{tab:constrains} summarizes the data adopted in our analysis: the luminosity of the Sun, its effective temperature, surface helium abundance, the surface metalliticy, the radius of the convective zone, the sound speed, as well as the measured neutrino flux components. We construct a $\chi^2$-test function, 
\begin{align}
    \chi^2 = \mathop{\Large\sum}_i\left(\frac{\mathcal{O}_i - \mathcal{P}_i}{\sigma_i}\right)^2\,,
\label{eq:chi2}
\end{align} 
where $\mathcal{O}_i$ denotes the central value for the $i$-th observable and $\mathcal{P}_i$ is the corresponding  model prediction. For each solar observable, the uncertainty $\sigma_i$ is adopted from the corresponding observational measurement, except for $Y_{\rm surface}$, $(Z/X)_{\rm surface}$, $R_{\rm cz}$, and $\delta\bar{c}_s^2/\bar{c}^2_s$, for which we adopt uncertainties three times larger than the  observational ones to avoid overconstraining the solar calibration with individual helioseismic and surface-composition diagnostics. This choice allows the different observables to contribute more evenly to the fit. We assume uncorrelated Gaussian uncertainties for all observables.

\begin{table}[t]
    \centering
    \renewcommand{\arraystretch}{1.6}
    \caption{Solar measurements included in our analysis: observed luminosity ($L$), surface temperature (also referred to as effective blackbody temperature, $T_{\rm eff}$), surface helium abundance ($Y_{\rm surface}$), surface metallicity ($(Z/X)_{\rm surface}$),  radius of the convective zone ($R_{\rm cz}$),  root-mean-square relative deviation of the sound-speed profile ($\bar{c}^2_s$), measured $pp$, $^7$Be, $^8$B, and CNO neutrino fluxes. For each quantity, we report the central value $\mathcal{O}_i$ and uncertainty $\sigma_i$ adopted in Eq.~\ref{eq:chi2}.
    }
    \begin{tabular}{lcc}
    \hline
        Parameter & Measurement & Ref.\\ \hline
        $L [L_\odot]$ & $1\pm 0.05$& \cite{IAU:2015fjh} \\
        $T_{\rm eff} [K] $ & $5772 \pm 65$ & \cite{IAU:2015fjh}\\
        $Y_{\rm surface}$ & $0.2485 \pm 0.01$ & \cite{Basu:2004zg}\\
        $(Z/X)_{\rm surface}$ & $0.023 \pm 0.003$ &\cite{Grevesse:1998bj}\\
        $R_{\rm cz}$ [$R_\odot$] & $0.713 \pm 0.003 $ & \cite{Basu:2004zg} \\
        $\delta\bar{c}_s^2/\bar{c}^2_s$ & $0 \pm 0.0006$ &\cite{Rhodes:1997}\\
        $\Phi_{\rm pp}$ [$\rm cm^{-2} s^{-1}$]& $(5.2\pm 0.1)\times 10^{10}$ &\cite{Gonzalez-Garcia:2023kva}\\
        $\Phi_{\rm ^7Be}$ [$\rm cm^{-2} s^{-1}$]& $ (4.93\pm0.10)\times 10^9$ &\cite{Gonzalez-Garcia:2023kva}\\
        $\Phi_{\rm ^8B}$ [$\rm cm^{-2} s^{-1}$]& $(5.941\pm0.024)\times10^{6}$ &\cite{Gonzalez-Garcia:2023kva}\\
        $\Phi_{\rm CNO}$ [$\rm cm^{-2} s^{-1}$]& $(6.7 \pm 1.2)\times 10^8$ &\cite{BOREXINO:2023ygs}\\ \hline
    \end{tabular}
    
    \label{tab:constrains}
\end{table}

For each combination of $\beta$ and $\alpha$ in Eq.~\ref{eq:parametrisation}, we minimize the $\chi^2$ in Eq.~\ref{eq:chi2} over the stellar parameters $\lbrace Y, Z_{\rm CNO}, Z_{\rm other}, \alpha_{\rm MLT}, f_{\rm ov}\rbrace$ for  solar ages in the range $[4.54, 4.68]$~Gyr, i.e., around the estimated solar age which we sample using $200$ non-uniformly spaced bins. We employ the Nelder-Mead simplex algorithm~\cite{Nelder:1965zz,Paxton:2013pj}, built following the solar-calibration routine provided in the \texttt{MESA test\_suite}, which we extend to include the additional energy-loss channel, the corresponding observational constraints, and the separation between initial composition of carbon, nitrogen and oxygen isotopes ($Z_{\rm CNO}$), and other elements ($Z_{\rm other}$): $Y \in [0.269,0.289]$, $Z_{\rm CNO}$ and $Z_{\rm other}$ in the range $[0, 0.02]$, $\alpha_{\rm MLT} \in [1.8, 2.0]$, and $f_{\rm ov} \in [0.014, 0.022]$.  
The variability ranges for such parameters are chosen ad hoc around the values required to obtain the standard evolution of the Sun and are sufficiently broad to accommodate the changes in the solar evolution induced by $\epsilon_{\rm{BSM}}$ (cf.~Eq.~\ref{eq:energy-conservation}). This approach sistematically captures  the impact of uncertainties in individual solar-model ingredients (such as opacities and metallicity) that may contribute to the solar modeling problem.

 We evaluate at least $100$ solar models for each point in the $(\beta, \alpha )$ parameter space in Fig.~\ref{fig:exclusion}. The solar models in agreement with  observations are the ones with  the minimum $\chi^2$. The minimization procedure is iterated until the relative change between successive estimates of the global minimum is $\lesssim 0.1\%$~\footnote{We have also verified that the global minimum of the $\chi^2$ function is found for the standard solar model, corresponding to $\alpha =0$, i.e.~for the null-hypothesis. For sufficiently small energy-loss normalizations, $\alpha < 10^{-2}$~erg~g$^{-1}$~s$^{-1}$, the minimized $\chi^2$ converges to the global minimum, independently of $\beta$.}. The number of evolved solar models can therefore exceed $100$ when additional iterations are required to achieve convergence. Overall, our analysis involves the evolution of  $\gtrsim 13,000$ solar models, each evaluated at $200$ steps in solar age, resulting in $\gtrsim 2.6 \times 10^6$ predictions of solar observables.

If $\epsilon_{\rm{BSM}}$ is too large,  the allowed uncertainty on  the stellar parameters becomes insufficient to reproduce the observed solar properties, allowing us to determine the extent of the exclusion  region in the parameter space spanned by $\beta$ and $\alpha$.
We stress that the exclusion region is determined  by current  solar observables: models that reproduce the  observables are not excluded if their past evolutionary history differs from the one computed for  $\epsilon_{\rm{BSM}} =0$.

\textit{Constraints on exotic energy losses.}---We compute the exclusion region in Fig.~\ref{fig:exclusion} (orange)  from the difference between the minimum $\chi^2$ for any given $(\beta, \alpha)$ and the global minimum, found for $\alpha =0$.
For two degrees of freedom (i.e.,  $\beta$ and $\alpha$) and with the stellar parameters profiled over, the $3 \sigma$ confidence contour is defined by $\Delta\chi^2=11.8$.  

The trend in Fig.~\ref{fig:exclusion} can be understood by examining  the consequences of varying  $\beta$ and $\alpha$, separately. For fixed $\beta$, as $\alpha$ increases, $\epsilon_{\rm BSM}$ (Eq.~\ref{eq:parametrisation})  increases. This leads to a reduction in $T_c$, accompanied by changes in the central density as the Sun adjusts to the modified energy budget (see Fig.~\ref{fig:evolution}).  
In turn, nuclear burning and the expected neutrino fluxes are affected (see also the Supplemental Material), as well as the helioseismic observables. Eventually, the resulting solar models cannot simultaneously reproduce the observational constraints, leading to their exclusion.
For fixed $\alpha$, an increase in $\beta$ enhances the temperature dependence of $\epsilon_{\rm BSM}$,   concentrating the energy loss  in the hottest regions of the solar core. This  not only changes the energy extracted from the Sun but also the evolution of the internal structure of the Sun 
(cf.~Fig.~\ref{fig:evolution}). The interplay between these two effects determines the shape of the exclusion region in Fig.~\ref{fig:exclusion}. Overall, the exclusion region in Fig.~\ref{fig:exclusion} suggests that exotic energy losses do not improve the overall agreement between solar models and observations, therefore  not providing a solution to the solar modeling problem.

Figure~\ref{fig:exclusion} can  be translated into limits on the luminosity emitted by the Sun in the form of exotic particles, yielding $L_{\rm BSM} \lesssim 0.10~L_\odot$ throughout the explored range $0\leq \beta \leq 10$. Further details on the limits on the solar luminosity due to exotic particles are provided in the Supplemental Material.

In order to understand whether the bounds in Fig.~\ref{fig:exclusion} may become more stringent thanks to  future improvements on the precision of solar neutrino measurements, we explore an optimistic scenario in which the uncertainties on the $pp$, $^7$Be, $^8$B, and CNO fluxes are reduced to $1\%$, $1\%$, $2\%$, and $10\%$, respectively; here, the choice of uncertainties reflects the  statistical and background limitations that affect the measurement of each component.
We find that an improvement in the  experimental uncertainties on all four neutrino flux components  would negligibly affect the  exclusion region in Fig.~\ref{fig:exclusion}. Further progress in this direction demands an improved  treatment of  evolution of the Sun in multi-dimensions as well as the characterization of the uncertainties entering the solar calibration~\cite{Basu:2016}.

{\it Constraints on axions and hidden photons.}---Among the best-motivated examples of particles beyond the Standard Model responsible for additional energy losses in stars are  fleebly-interacting particles, such as axions, axion-like particles, and hidden (or dark) photons. Axions are produced in stellar plasmas through Primakoff conversion of thermal photons in the electromagnetic field of charged particles~\cite{ChadhaDay:2022,PDG:2024,Carenza:2024ehj}.  Hidden photons  can have  two  polarizations (transversal and longitudinal) and are produced through kinetic mixing with photons~\cite{Agrawal:2021dbo,Berlin:2024pzi,Caputo:2025avc}. Longitudinal hidden photons are expected to be resonantly generated when their in-medium properties match those of plasma photon modes~\cite{Redondo:2013lna}. These exotic particles lead to temperature-dependent energy loss from the stellar core: the Primakoff energy loss rate is $\varepsilon \propto T^7$~\cite{Raffelt:1985nk}, while the resonant longitudinal hidden-photon production scales as $\varepsilon \propto T$ in the Sun~\cite{Redondo:2013lna}.

As summarized in Fig.~\ref{fig:exclusion}, for 
$\beta=7$, our parameterization approximately mimics the energy loss from Primakoff axion production and leads to $\alpha \lesssim 0.05$~erg~g$^{-1}$~s$^{-1}$ (or $L_{\rm BSM} \lesssim 0.04~L_\odot$); and for $\beta=1$ it resembles resonant production of longitudal hidden photons and yields $\alpha \lesssim 0.2$~erg~g$^{-1}$~s$^{-1}$ (or $L_{\rm BSM} \lesssim 0.10~L_\odot$). 
These results offer an independent and complementary approach to
ongoing searches for fleebly-interacting particles that are broadly divided into direct and indirect. 

Direct searches include helioscopes, which aim to detect axions produced in the Sun, and haloscopes that look for axions from the Galactic dark matter halo~\cite{Sikivie:1983ip,CAST:2017uph,ADMX:2018gho}. Direct searches for solar axions from the CERN Solar Axion Telescope (CAST)~\cite{CAST:2017uph} and solar hidden photons from XENON10~\cite{An:2013yua} provide  constraints $70$ and $2$ times more stringent on $\alpha$ than our analysis, respectively.

Indirect searches exploit the impact of new particles on stellar evolution and cooling. These include constraints from the evolution of main-sequence stars, the tip of the red-giant branch, white-dwarf and neutron-star cooling, as well as the evolution of massive stars, see e.g.~Refs.~\cite{Raffelt:1996wa,Raffelt:1999tx,DiLuzio:2021ysg,Carenza:2024ehj}. Primakoff axion production scales approximately as $T^7$, hence axions may be prone to greater energy loss in environments hotter than the Sun, such as red-giant stars, horizontal-branch stars, and core-collapse supernovae,  leading to complementary and potentially more stringent constraints~\cite{Carenza:2024ehj}. For hidden photons, the strongest astrophysical constraints  depend  on the particle mass and polarization mode. In particular, solar observations provide competitive constraints on longitudinal hidden-photon production at low masses, typically below $0.1$--$1$~keV~\cite{Redondo:2013lna}, whereas horizontal-branch and red-giant stars can provide stronger bounds with transverse-mode production becoming particularly relevant at higher masses, around and above the keV scale~\cite{An:2014twa,Dolan:2023cjs}. Core-collapse supernovae provide additional constraints at higher masses, reaching the MeV range~\cite{Caputo:2025aac}.  We refer the interested reader to the Supplemental Material for a detailed comparison of our results with existing bounds on axions and hidden photons.

Solar bounds for axions and hidden photons were computed in  Ref.~\cite{Vinyoles:2015aba}  combining helioseismic observables and  the measured $^7$Be and $^8$B neutrino fluxes. Overall, our constraints on $\alpha$ are weaker (yet more robust) than the ones in Ref.~\cite{Vinyoles:2015aba} by a factor $2.5$ for Primakoff axion production and $5$ for longitudinal hidden-photon emission. 
However,  Ref.~\cite{Vinyoles:2015aba} only used  two  values of $\beta$ in our parameter space (cf.~Supplemental Material). 
Our analysis differs from the one in Ref.~\cite{Vinyoles:2015aba} in several aspects. First, we evolve solar models with {\tt MESA}; 
Ref.~\cite{Vinyoles:2015aba} employs GARSTEC (GARching STEllar Code)~\cite{Weiss:2008} instead. Second, we consider an end-to-end treatment of uncertainties on the initial chemical composition, convection parameters, and solar age, and profile over them, whereas  Refs.~\cite{Vinyoles:2015aba,Vinyoles:2015khy} account for theoretical uncertainties through linearized propagation of uncertainties in stellar parameters; this approach does not capture the non-linear feedback of such uncertainties on the stellar evolution nor the interplay with the exotic energy losses. 
Finally, our analysis includes a broader set of  solar observables, including the \textit{pp} and CNO neutrino fluxes.

\textit{Conclusions.}---Non-standard energy loss  mechanisms in the Sun can affect its evolution.  
We evolve $\gtrsim 13,000$ spherically-symmetric solar models using {\tt MESA}, including an additional energy-loss channel ($\propto \alpha T^{\beta}$, with $0 \lesssim  \beta \lesssim 10$ and $0.01\lesssim \alpha\lesssim 1$~erg~g$^{-1}$~s$^{-1}$) and  considering uncertainties in the  initial chemical composition of the Sun, convection, and the age of the Sun.  This is the first investigation of non-standard energy losses spanning a wide parameter space and consistently  tracking   changes in stellar structure and nuclear burning associated to physics beyond the Standard Model.  

By comparing the outcome of the solar models with the latest measurements of the $pp$, $^7$Be, $^8$B, and CNO neutrino fluxes, helioseismic data as well as global solar observables,  our analysis establishes a robust upper bound on energy losses of $\alpha \lesssim 0.25$~erg~g$^{-1}$~s$^{-1}$ (or, equivalently, $L_{\rm BSM} \lesssim 0.10 L_\odot$).
The approach presented in this work highlights that the temperature sensitivity of $^7$Be and $^8$B neutrinos, together with the dependence of CNO neutrinos on both temperature and chemical composition, is crucial to  constrain departures from the standard solar evolution. Notably, our results suggest that exotic energy losses do not alleviate the solar modeling problem.

We infer novel solar bounds on Primakoff axion production ($\alpha \lesssim 0.05$~erg~g$^{-1}$~s$^{-1}$ for $\beta =7$) and hidden-photon emission ($\alpha \lesssim 0.2$~erg~g$^{-1}$~s$^{-1}$ for $\beta =1$).  
These results complement  direct searches led by CAST and XENON10.

The  framework outlined here applies to a broad class of exotic energy-loss mechanisms and can be extended to more general energy-loss mechanisms, for example  exotic scenarios depending on the baryon density or electron fraction. 
By combining solar observations with advanced solar evolution modeling, this work established a novel and powerful approach for probing non-standard physics in any stellar environment. 

\textit{Acknowledgements.}---We thank Georg Raffelt for insightful comments on the manuscript. This research was supported by the Elite Research Prize from the Danish Ministry of Higher Education and Science (Project No.~3142-00074B),  Villum Fonden (Project No.~VIL82948), and in part by grant NSF PHY-2309135 to the Kavli Institute for Theoretical Physics (KITP).  The Tycho supercomputer hosted at the SCIENCE HPC center at the University of Copenhagen was used to support the numerical simulations presented in this work.

\bibliographystyle{apsrev4-2}
\bibliography{bibliography}

@book{Raffelt:1996wa,
  author    = {Raffelt, Georg G.},
  title     = {Stars as Laboratories for Fundamental Physics:     The Astrophysics of Neutrinos, Axions, and Other Weakly Interacting Particles},
  publisher = {University of Chicago Press},
  address   = {Chicago},
  year      = {1996},
  isbn      = {978-0-226-70272-8}
}

@article{Vinyoles:2015aba,
    author = "Vinyoles, N{\'u}ria and Serenelli, Aldo and Villante, Francesco L. and Basu, Sarbani and Redondo, Javier and Isern, Jordi",
    title = "{New axion and hidden photon constraints from a solar data global fit}",
    eprint = "1501.01639",
    archivePrefix = "arXiv",
    primaryClass = "astro-ph.SR",
    doi = "10.1088/1475-7516/2015/10/015",
    journal = "JCAP",
    volume = "10",
    pages = "015",
    year = "2015"
}

@article{Schlattl:1998fz,
    author = "Schlattl, H. and Weiss, A. and Raffelt, G.",
    title = "{Helioseismological constraint on solar axion emission}",
    eprint = "hep-ph/9807476",
    archivePrefix = "arXiv",
    doi = "10.1016/S0927-6505(98)00063-2",
    journal = "Astropart. Phys.",
    volume = "10",
    pages = "353--359",
    year = "1999"
}

@article{Chen:2024ouz,
    author = "Chen, X. and others",
    title = "{Enhanced S-Factor for the N14(p,{\ensuremath{\gamma}})15O Reaction and Its Impact on the Solar Composition Problem}",
    eprint = "2410.16086",
    archivePrefix = "arXiv",
    primaryClass = "nucl-ex",
    doi = "10.1103/q756-hzmt",
    journal = "Phys. Rev. Lett.",
    volume = "135",
    number = "23",
    pages = "232701",
    year = "2025"
}

@article{Basu:2007fp,
    author = "Basu, Sarbani and Antia, H. M.",
    title = "{Helioseismology and Solar Abundances}",
    eprint = "0711.4590",
    archivePrefix = "arXiv",
    primaryClass = "astro-ph",
    doi = "10.1016/j.physrep.2007.12.002",
    journal = "Phys. Rept.",
    volume = "457",
    pages = "217--283",
    year = "2008"
}

@article{Boothroyd:2002qw,
    author = "Boothroyd, Arnold I. and Sackmann, I. Juliana",
    title = "{Our sun. 4. The standard model and helioseismology: consequences of uncertainties in input physics and in observed solar parameters}",
    eprint = "astro-ph/0210127",
    archivePrefix = "arXiv",
    reportNumber = "CITA-2002-35",
    doi = "10.1086/345407",
    journal = "Astrophys. J.",
    volume = "583",
    pages = "1004--1023",
    year = "2003"
}

@article{Asplund:2009fu,
    author = "Asplund, Martin and Grevesse, Nicolas and Sauval, A. Jacques and Scott, Pat",
    title = "{The chemical composition of the Sun}",
    eprint = "0909.0948",
    archivePrefix = "arXiv",
    primaryClass = "astro-ph.SR",
    doi = "10.1146/annurev.astro.46.060407.145222",
    journal = "Ann. Rev. Astron. Astrophys.",
    volume = "47",
    pages = "481--522",
    year = "2009"
}

@ARTICLE{2021A&A...653A.141A,
       author = {{Asplund}, M. and {Amarsi}, A.~M. and {Grevesse}, N.},
        title = "{The chemical make-up of the Sun: A 2020 vision}",
      journal = {Astron. Astrophys.},
         year = 2021,
        month = sep,
       volume = {653},
          eid = {A141},
        pages = {A141},
          doi = {10.1051/0004-6361/202140445},
archivePrefix = {arXiv},
       eprint = {2105.01661},
 primaryClass = {astro-ph.SR},
       adsurl = {https://scixplorer.org/abs/2021A&A...653A.141A}
}

@article{Buldgen:2024aak,
    author = "Buldgen, G. and others",
    title = "{In-depth analysis of solar models with high-metallicity abundances and updated opacity tables}",
    eprint = "2404.10478",
    archivePrefix = "arXiv",
    primaryClass = "astro-ph.SR",
    doi = "10.1051/0004-6361/202348312",
    journal = "Astron. Astrophys.",
    volume = "686",
    pages = "A108",
    year = "2024"
}

@article{Christensen-Dalsgaard:2008idi,
    author = "Christensen-Dalsgaard, Jorgen and Di Mauro, Maria Pia and Houdek, Gunter and Pijpers, Frank",
    title = "{On the opacity change required to compensate for the revised solar composition}",
    eprint = "0811.1001",
    archivePrefix = "arXiv",
    primaryClass = "astro-ph",
    doi = "10.1051/0004-6361:200810170",
    journal = "Astron. Astrophys.",
    volume = "494",
    pages = "205",
    year = "2009"
}

@article{Gonzalez-Garcia:2023kva,
    author = "Gonzalez-Garcia, M. C. and Maltoni, Michele and Pinheiro, Jo{\~a}o Paulo and Serenelli, Aldo M.",
    title = "{Status of direct determination of solar neutrino fluxes after Borexino}",
    eprint = "2311.16226",
    archivePrefix = "arXiv",
    primaryClass = "hep-ph",
    reportNumber = "IFT-UAM/CSIC-24-3, YITP-SB-23-39",
    doi = "10.1007/JHEP02(2024)064",
    journal = "JHEP",
    volume = "02",
    pages = "064",
    year = "2024"
}

@article{Bahcall:1996vj,
    author = "Bahcall, John N. and Ulmer, Andrew",
    title = "{The Temperature dependence of solar neutrino fluxes}",
    eprint = "astro-ph/9602012",
    archivePrefix = "arXiv",
    reportNumber = "IASSNS-AST-96-8",
    doi = "10.1103/PhysRevD.53.4202",
    journal = "Phys. Rev. D",
    volume = "53",
    pages = "4202--4210",
    year = "1996"
}

@article{Paxton:2011,
  author        = {Paxton, Bill and others},
  title         = {Modules for Experiments in Stellar Astrophysics (MESA)},
  journal       = {Astrophys. J. Suppl.},
  volume        = {192},
  pages         = {3},
  year          = {2011},
  doi           = {10.1088/0067-0049/192/1/3},
  archivePrefix = {arXiv},
  eprint        = {1009.1622}
}

@article{Paxton:2019,
  author        = {Paxton, Bill and others},
  title         = {Modules for Experiments in Stellar Astrophysics (MESA):
                   Pulsating Variable Stars, Rotation, Convective
                   Boundaries, and Energy Conservation},
  journal       = {Astrophys. J. Suppl.},
  volume        = {243},
  pages         = {10},
  year          = {2019},
  doi           = {10.3847/1538-4365/ab2241},
  archivePrefix = {arXiv},
  eprint        = {1903.01426}
}

@article{Paxton:2013pj,
    author = "Paxton, Bill and others",
    title = "{Modules for Experiments in Stellar Astrophysics (MESA): Planets, Oscillations, Rotation, and Massive Stars}",
    eprint = "1301.0319",
    archivePrefix = "arXiv",
    primaryClass = "astro-ph.SR",
    doi = "10.1088/0067-0049/208/1/4",
    journal = "Astrophys. J. Suppl.",
    volume = "208",
    pages = "4",
    year = "2013"
}

@article{Paxton:2015jva,
    author = "Paxton, Bill and others",
    title = "{Modules for Experiments in Stellar Astrophysics (MESA): Binaries, Pulsations, and Explosions}",
    eprint = "1506.03146",
    archivePrefix = "arXiv",
    primaryClass = "astro-ph.SR",
    doi = "10.1088/0067-0049/220/1/15",
    journal = "Astrophys. J. Suppl.",
    volume = "220",
    number = "1",
    pages = "15",
    year = "2015"
}

@article{MESA:2022zpy,
    author = "Jermyn, Adam S. and others",
    collaboration = "MESA",
    title = "{Modules for Experiments in Stellar Astrophysics (MESA): Time-dependent Convection, Energy Conservation, Automatic Differentiation, and Infrastructure}",
    eprint = "2208.03651",
    archivePrefix = "arXiv",
    primaryClass = "astro-ph.SR",
    doi = "10.3847/1538-4365/acae8d",
    journal = "Astrophys. J. Suppl.",
    volume = "265",
    number = "1",
    pages = "15",
    year = "2023"
}

@article{Henyey:1965,
  author  = {Henyey, Louis and Vardya, M. S. and Bodenheimer, Peter},
  title   = {Studies in Stellar Evolution. III. The Calculation
             of Model Envelopes},
  journal = {Astrophys. J.},
  volume  = {142},
  pages   = {841},
  year    = {1965},
  doi     = {10.1086/148357}
}

@article{Herwig:2000sq,
    author = "Herwig, Falk",
    title = "{The evolution of agb stars with convective overshoot}",
    eprint = "astro-ph/0007139",
    archivePrefix = "arXiv",
    journal = "Astron. Astrophys.",
    volume = "360",
    pages = "952--968",
    year = "2000"
}

@article{BOREXINO:2023ygs,
    author = "Basilico, D. and others",
    collaboration = "BOREXINO",
    title = "{Final results of Borexino on CNO solar neutrinos}",
    eprint = "2307.14636",
    archivePrefix = "arXiv",
    primaryClass = "hep-ex",
    doi = "10.1103/PhysRevD.108.102005",
    journal = "Phys. Rev. D",
    volume = "108",
    number = "10",
    pages = "102005",
    year = "2023"
}

@article{Grevesse:1998bj,
    author = "Grevesse, N. and Sauval, A. J.",
    title = "{Standard Solar Composition}",
    doi = "10.1023/A:1005161325181",
    journal = "Space Sci. Rev.",
    volume = "85",
    pages = "161--174",
    year = "1998"
}

@article{Basu:1997,
  author  = {Basu, Sarbani and Antia, H. M.},
  title   = {Seismic Measurement of the Depth of the Solar
             Convection Zone},
  journal = {Mon. Not. Roy. Astron. Soc.},
  volume  = {287},
  pages   = {189--198},
  year    = {1997},
  doi     = {10.1093/mnras/287.1.189}
}

@article{Basu:2004zg,
    author = "Basu, Sarbani and Antia, H. M.",
    title = "{Constraining solar abundances using helioseismology}",
    eprint = "astro-ph/0403485",
    archivePrefix = "arXiv",
    doi = "10.1086/421110",
    journal = "Astrophys. J. Lett.",
    volume = "606",
    pages = "L85",
    year = "2004"
}

@article{Schwarz:2015lqa,
    author = {Schwarz, Matthias and Knabbe, Ernst-Axel and Lindner, Axel and Redondo, Javier and Ringwald, Andreas and Schneide, Magnus and Susol, Jaroslaw and Wiedemann, G{\"u}nter},
    title = "{Results from the Solar Hidden Photon Search (SHIPS)}",
    eprint = "1502.04490",
    archivePrefix = "arXiv",
    primaryClass = "hep-ph",
    reportNumber = "DESY-15-030",
    doi = "10.1088/1475-7516/2015/08/011",
    journal = "JCAP",
    volume = "08",
    pages = "011",
    year = "2015"
}

@article{CAST:2017uph,
    author = "Anastassopoulos, V. and others",
    collaboration = "CAST",
    title = "{New CAST Limit on the Axion-Photon Interaction}",
    eprint = "1705.02290",
    archivePrefix = "arXiv",
    primaryClass = "hep-ex",
    doi = "10.1038/nphys4109",
    journal = "Nature Phys.",
    volume = "13",
    pages = "584--590",
    year = "2017"
}

@article{Rhodes:1997,
  author  = {Rhodes, E. J., Jr. and Kosovichev, A. G. and
             Schou, J. and Scherrer, P. H. and Reiter, J.},
  title   = {Measurements of Frequencies of Solar Oscillations
             from the MDI Medium-l Program},
  journal = {Sol. Phys.},
  volume  = {175},
  pages   = {287--310},
  year    = {1997},
  doi     = {10.1023/A:1004963425123}
}

@article{Raffelt:1999tx,
  author = {Raffelt, Georg G.},
  title = {Particle Physics from Stars},
  journal = {Ann. Rev. Nucl. Part. Sci.},
  volume = {49},
  pages = {163--216},
  year = {1999},
  doi = {10.1146/annurev.nucl.49.1.163}
}

@article{Sikivie:1983ip,
    author = "Sikivie, P.",
    editor = "Srednicki, M. A.",
    title = "{Experimental Tests of the Invisible Axion}",
    reportNumber = "PRINT-83-0597 (FLORIDA), UF-TP-83-13",
    doi = "10.1103/PhysRevLett.51.1415",
    journal = "Phys. Rev. Lett.",
    volume = "51",
    pages = "1415--1417",
    year = "1983",
    note = "[Erratum: Phys.Rev.Lett. 52, 695 (1984)]"
}

@article{Carenza:2024ehj,
    author = "Carenza, Pierluca and Giannotti, Maurizio and Isern, Jordi and Mirizzi, Alessandro and Straniero, Oscar",
    title = "{Axion astrophysics}",
    eprint = "2411.02492",
    archivePrefix = "arXiv",
    primaryClass = "hep-ph",
    reportNumber = "BARI-TH/66-24",
    doi = "10.1016/j.physrep.2025.02.002",
    journal = "Phys. Rept.",
    volume = "1117",
    pages = "1--102",
    year = "2025"
}

@article{ADMX:2018gho,
    author = "Du, N. and others",
    collaboration = "ADMX",
    title = "{A Search for Invisible Axion Dark Matter with the Axion Dark Matter Experiment}",
    eprint = "1804.05750",
    archivePrefix = "arXiv",
    primaryClass = "hep-ex",
    reportNumber = "FERMILAB-PUB-18-101-AD-AE",
    doi = "10.1103/PhysRevLett.120.151301",
    journal = "Phys. Rev. Lett.",
    volume = "120",
    number = "15",
    pages = "151301",
    year = "2018"
}

@article{IAU:2015fjh,
    author = "Mamajek, E. E. and others",
    collaboration = "{IAU Inter-Division A-G Working Group on Nominal Units for Stellar \& Planetary Astronomy}",
    title = "{IAU 2015 Resolution B3 on Recommended Nominal Conversion Constants for Selected Solar and Planetary Properties}",
    eprint = "1510.07674",
    archivePrefix = "arXiv",
    primaryClass = "astro-ph.SR",
    month = "10",
    year = "2015",
    journal ="",
}

@article{Raffelt:1985nk,
    author = "Raffelt, Georg G.",
    title = "{ASTROPHYSICAL AXION BOUNDS DIMINISHED BY SCREENING EFFECTS}",
    reportNumber = "MPI-PAE-PTH-51-85",
    doi = "10.1103/PhysRevD.33.897",
    journal = "Phys. Rev. D",
    volume = "33",
    pages = "897",
    year = "1986"
}

@article{DiLuzio:2021ysg,
    author = "Di Luzio, Luca and Fedele, Marco and Giannotti, Maurizio and Mescia, Federico and Nardi, Enrico",
    title = "{Stellar evolution confronts axion models}",
    eprint = "2109.10368",
    archivePrefix = "arXiv",
    primaryClass = "hep-ph",
    reportNumber = "DESY-21-141, TTP21-030, P3H-21-062",
    doi = "10.1088/1475-7516/2022/02/035",
    journal = "JCAP",
    volume = "02",
    number = "02",
    pages = "035",
    year = "2022"
}

@article{Gondolo:2008dd,
    author = "Gondolo, Paolo and Raffelt, Georg G.",
    title = "{Solar neutrino limit on axions and keV-mass bosons}",
    eprint = "0807.2926",
    archivePrefix = "arXiv",
    primaryClass = "astro-ph",
    reportNumber = "MPP-2008-69",
    doi = "10.1103/PhysRevD.79.107301",
    journal = "Phys. Rev. D",
    volume = "79",
    pages = "107301",
    year = "2009"
}

@article{Redondo:2013lna,
    author = "Redondo, Javier and Raffelt, Georg G.",
    title = "{Solar constraints on hidden photons re-visited}",
    eprint = "1305.2920",
    archivePrefix = "arXiv",
    primaryClass = "hep-ph",
    reportNumber = "MPP-2013-110",
    doi = "10.1088/1475-7516/2013/08/034",
    journal = "JCAP",
    volume = "08",
    pages = "034",
    year = "2013"
}

@article{Borexino:2017rsf,
    author = "Agostini, M. and others",
    collaboration = "Borexino",
    title = "{First Simultaneous Precision Spectroscopy of $pp$, $^7$Be, and $pep$ Solar Neutrinos with Borexino Phase-II}",
    eprint = "1707.09279",
    archivePrefix = "arXiv",
    primaryClass = "hep-ex",
    doi = "10.1103/PhysRevD.100.082004",
    journal = "Phys. Rev. D",
    volume = "100",
    number = "8",
    pages = "082004",
    year = "2019"
}

@article{An:2013yua,
    author = "An, Haipeng and Pospelov, Maxim and Pradler, Josef",
    title = "{Dark Matter Detectors as Dark Photon Helioscopes}",
    eprint = "1304.3461",
    archivePrefix = "arXiv",
    primaryClass = "hep-ph",
    reportNumber = "PI-PARTPHYS-324",
    doi = "10.1103/PhysRevLett.111.041302",
    journal = "Phys. Rev. Lett.",
    volume = "111",
    pages = "041302",
    year = "2013"
}

@article{Borexino:2017uhp,
    author = "Agostini, M. and others",
    collaboration = "Borexino",
    title = {{Improved measurement of $^8$B solar neutrinos with $1.5\,\mathrm{kt}\,\mathrm{yr}$ of Borexino exposure}},
    eprint = "1709.00756",
    archivePrefix = "arXiv",
    primaryClass = "hep-ex",
    doi = "10.1103/PhysRevD.101.062001",
    journal = "Phys. Rev. D",
    volume = "101",
    number = "6",
    pages = "062001",
    year = "2020"
}

@article{Agrawal:2021dbo,
    author = "Agrawal, Prateek and others",
    title = "{Feebly-interacting particles: FIPs 2020 workshop report}",
    eprint = "2102.12143",
    archivePrefix = "arXiv",
    primaryClass = "hep-ph",
    doi = "10.1140/epjc/s10052-021-09703-7",
    journal = "Eur. Phys. J. C",
    volume = "81",
    number = "11",
    pages = "1015",
    year = "2021"
}

@article{Berlin:2024pzi,
    author = "Berlin, Asher and Kahn, Yonatan",
    title = "{New Technologies for Axion and Dark Photon Searches}",
    eprint = "2412.08704",
    archivePrefix = "arXiv",
    primaryClass = "hep-ph",
    reportNumber = "FERMILAB-PUB-24-0933-T",
    doi = "10.1146/annurev-nucl-121423-101015",
    journal = "Ann. Rev. Nucl. Part. Sci.",
    volume = "75",
    number = "1",
    pages = "83--108",
    year = "2025"
}

@article{BOREXINO:2022abl,
    author = "Appel, S. and others",
    collaboration = "BOREXINO",
    title = "{Improved Measurement of Solar Neutrinos from the Carbon-Nitrogen-Oxygen Cycle by Borexino and Its Implications for the Standard Solar Model}",
    eprint = "2205.15975",
    archivePrefix = "arXiv",
    primaryClass = "hep-ex",
    doi = "10.1103/PhysRevLett.129.252701",
    journal = "Phys. Rev. Lett.",
    volume = "129",
    number = "25",
    pages = "252701",
    year = "2022"
}

@article{Basu:2016,
  author  = {Basu, Sarbani},
  title   = {Global Seismology of the Sun},
  journal = {Living Rev. Sol. Phys.},
  volume  = {13},
  pages   = {2},
  year    = {2016},
  doi     = {10.1007/s41116-016-0003-4}
}

@article{Nelder:1965zz,
    author = "Nelder, J. A. and Mead, R.",
    title = "{A Simplex Method for Function Minimization}",
    doi = "10.1093/comjnl/7.4.308",
    journal = "Comput. J.",
    volume = "7",
    pages = "308--313",
    year = "1965"
}

@article{Caputo:2025aac,
    author = "Caputo, Andrea and Janka, Hans-Thomas and Raffelt, Georg G. and Yun, Seokhoon",
    title = "{Cooling the Shock: New Supernova Constraints on Dark Photons}",
    eprint = "2502.01731",
    archivePrefix = "arXiv",
    primaryClass = "hep-ph",
    reportNumber = "CERN-TH-2025-020, CTPU-PTC-25-05",
    doi = "10.1103/PhysRevLett.134.151002",
    journal = "Phys. Rev. Lett.",
    volume = "134",
    number = "15",
    pages = "151002",
    year = "2025"
}

@article{Caputo:2025avc,
    author = "Caputo, Andrea and Park, Jaeyoung and Yun, Seokhoon",
    title = "{Heavy dark photon handbook: Cosmological and astrophysical bounds}",
    eprint = "2511.15785",
    archivePrefix = "arXiv",
    primaryClass = "hep-ph",
    reportNumber = "CTPU-PTC-25-42",
    doi = "10.1103/m6zs-jxxp",
    journal = "Phys. Rev. D",
    volume = "113",
    number = "7",
    pages = "075001",
    year = "2026"
}

@article{An:2014twa,
    author = "An, Haipeng and Pospelov, Maxim and Pradler, Josef and Ritz, Adam",
    title = "{Direct Detection Constraints on Dark Photon Dark Matter}",
    eprint = "1412.8378",
    archivePrefix = "arXiv",
    primaryClass = "hep-ph",
    reportNumber = "CALT-TH-2014-173",
    doi = "10.1016/j.physletb.2015.06.018",
    journal = "Phys. Lett. B",
    volume = "747",
    pages = "331--338",
    year = "2015"
}

@article{Dolan:2023cjs,
    author = "Dolan, Matthew J. and Hiskens, Frederick J. and Volkas, Raymond R.",
    title = "{Constraining dark photons with self-consistent simulations of globular cluster stars}",
    eprint = "2306.13335",
    archivePrefix = "arXiv",
    primaryClass = "hep-ph",
    doi = "10.1088/1475-7516/2024/05/099",
    journal = "JCAP",
    volume = "05",
    pages = "099",
    year = "2024"
}

@article{Peccei:1977hh,
    author = "Peccei, R. D. and Quinn, Helen R.",
    title = "{CP Conservation in the Presence of Instantons}",
    reportNumber = "ITP-568-STANFORD",
    doi = "10.1103/PhysRevLett.38.1440",
    journal = "Phys. Rev. Lett.",
    volume = "38",
    pages = "1440--1443",
    year = "1977"
}

@article{Weinberg:1977ma,
    author = "Weinberg, Steven",
    title = "{A New Light Boson?}",
    reportNumber = "HUTP-77/A074",
    doi = "10.1103/PhysRevLett.40.223",
    journal = "Phys. Rev. Lett.",
    volume = "40",
    pages = "223--226",
    year = "1978"
}

@article{Wilczek:1977pj,
    author = "Wilczek, Frank",
    title = "{Problem of Strong  $P$  and  $T$  Invariance in the Presence of Instantons}",
    reportNumber = "Print-77-0939 (COLUMBIA)",
    doi = "10.1103/PhysRevLett.40.279",
    journal = "Phys. Rev. Lett.",
    volume = "40",
    pages = "279--282",
    year = "1978"
}

@article{Okun:1982xi,
    author = "Okun, L. B.",
    title = "{LIMITS OF ELECTRODYNAMICS: PARAPHOTONS?}",
    reportNumber = "ITEP-48-1982",
    journal = "Sov. Phys. JETP",
    volume = "56",
    pages = "502",
    year = "1982"
}

@article{Holdom:1985ag,
    author = "Holdom, Bob",
    title = "{Two U(1)'s and Epsilon Charge Shifts}",
    reportNumber = "UTPT-85-30",
    doi = "10.1016/0370-2693(86)91377-8",
    journal = "Phys. Lett. B",
    volume = "166",
    pages = "196--198",
    year = "1986"
}

@article{ChadhaDay:2022,
    author = "Chadha-Day, Francesca and Ellis, John and Marsh, David J. E.",
    title = "{Axion dark matter: What is it and why now?}",
    journal = "Sci. Adv.",
    volume = "8",
    number = "8",
    pages = "eabj3618",
    year = "2022",
    doi = "10.1126/sciadv.abj3618"
}

@article{PDG:2024,
    author = "Navas, S. and others",
    collaboration = "Particle Data Group",
    title = "{Review of Particle Physics}",
    journal = "Phys. Rev. D",
    volume = "110",
    number = "3",
    pages = "030001",
    year = "2024",
    doi = "10.1103/PhysRevD.110.030001"
}

@article{Weiss:2008,
    author  = {Weiss, Achim and Schlattl, Helmut},
    title   = {{GARSTEC}---the {Garching} Stellar Evolution Code},
    journal = {Astrophysics and Space Science},
    volume  = {316},
    pages   = {99--106},
    year    = {2008},
    doi     = {10.1007/s10509-007-9606-5}
}

@article{Vinyoles:2015khy,
    author = "Vinyoles, N{\'u}ria and Vogel, Hendrik",
    title = "{Minicharged Particles from the Sun: A Cutting-Edge Bound}",
    eprint = "1511.01122",
    archivePrefix = "arXiv",
    primaryClass = "hep-ph",
    reportNumber = "MPP-2015-199",
    doi = "10.1088/1475-7516/2016/03/002",
    journal = "JCAP",
    volume = "03",
    pages = "002",
    year = "2016"
}

@article{Amarsi:2026,
      title={Solar chemical composition}, 
      author={A. M. Amarsi and N. Grevesse},
      year={2026},
      eprint={2608.23155},
      archivePrefix={arXiv},
      primaryClass={astro-ph.SR},
      url={https://arxiv.org/abs/2608.23155}, 
      journal =""
}

@article{Buldgen_2025,
       author = {{Buldgen}, Ga{\"e}l and {Pain}, Jean-Christophe and {Coss{\'e}}, Philippe and {Blancard}, Christophe and {Gilleron}, Franck and {Pradhan}, Anil K. and {Fontes}, Christopher J. and {Colgan}, James and {Noels}, Arlette and {Christensen-Dalsgaard}, J{\o}rgen and et al.},
        title = "{Helioseismic inference of the solar radiative opacity}",
      journal = {Nature Communications},
         year = 2025,
        month = jan,
       volume = {16},
       number = {1},
          eid = {693},
        pages = {693},
          doi = {10.1038/s41467-024-54793-y},
archivePrefix = {arXiv},
       eprint = {2504.06891},
 primaryClass = {astro-ph.SR},
       adsurl = {https://scixplorer.org/abs/2025NatCo..16..693B}
}

@ARTICLE{Buldgen:2024,
       author = {{Buldgen}, G. and {Noels}, A. and {Scuflaire}, R. and {Amarsi}, A.~M. and {Grevesse}, N. and {Eggenberger}, P. and {Colgan}, J. and {Fontes}, C.~J. and {Baturin}, V.~A. and {Oreshina}, A.~V. and et al.},
        title = "{In-depth analysis of solar models with high-metallicity abundances and updated opacity tables}",
      journal = {\aap},
         year = 2024,
        month = jun,
       volume = {686},
          eid = {A108},
        pages = {A108},
          doi = {10.1051/0004-6361/202348312},
archivePrefix = {arXiv},
       eprint = {2404.10478},
 primaryClass = {astro-ph.SR},
       adsurl = {https://scixplorer.org/abs/2024A&A...686A.108B}
}

\clearpage
\onecolumngrid
\appendix

\newpage
\clearpage

\setcounter{equation}{0}
\setcounter{figure}{0}
\setcounter{table}{0}
\setcounter{page}{1}
\makeatletter
\renewcommand{\theequation}{S\arabic{equation}}
\renewcommand{\thefigure}{S\arabic{figure}}
\renewcommand{\thetable}{S\arabic{table}}
\renewcommand{\thepage}{S\arabic{page}}

\begin{center}
\textbf{\large Supplemental Material \\[0.8ex]
{Neutrinos Probe Exotic Energy Losses from the Sun}}
\end{center}

In this Supplemental Material, we present our constraints in terms of the luminosity emitted in exotic particles  and we provide additional details on 
the dependence of the neutrino fluxes on  exotic energy losses. We also discuss how our  parameterization can be  mapped onto specific beyond-the-Standard-Model scenarios, focusing on Primakoff axion production and resonant production of longitudinal hidden photons.

\section{A.~Limits on the solar luminosity due to exotic energy losses\label{sec:BSM-luminoisty}}

Our constraints on the local energy-rate loss per unit mass  (Eq.~\ref{eq:parametrisation}) can  be translated into bounds on the  luminosity emitted by the Sun in non-standard channels: 
\begin{equation}
    L_{\rm BSM} = {\displaystyle\int}^{ M_\odot}_0 {\rm d \varepsilon_{BSM}}{\rm d}m = \displaystyle\int^{M_\odot}_0 \alpha\left(\frac{T (m)}{10^7 K}\right)^\beta {\rm d}m \, ,
\end{equation}
where $M_\odot$ is the solar mass and $m$ represents the mass enclosed in each mass cell in \texttt{MESA}.

\begin{figure}[h!]
    \centering
    \includegraphics[width=0.5\linewidth]{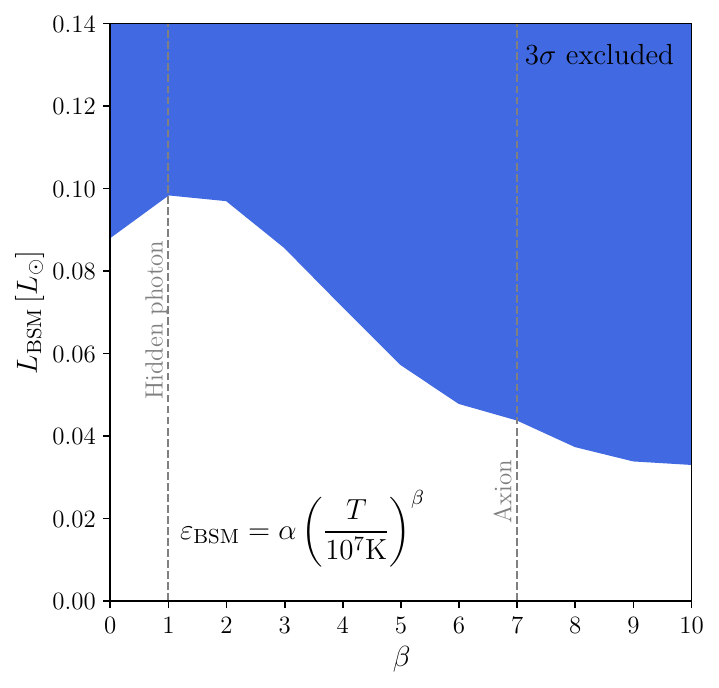}
    \caption{Same as Fig.~\ref{fig:exclusion}, but the exclusion bounds are expressed in terms of the  luminosity of exotic particles from the Sun,  in units of solar photon luminosity ($L_\odot$). These bounds refer to the current luminosity carried  by exotic particles and do not  quantify the corresponding reduction in photon luminosity, since the solar structure and nuclear energy generation readjust in response to the additional energy-loss channel.
    }
    \label{fig:luminosity}
\end{figure}

Figure~\ref{fig:luminosity} shows the limits on the  luminosity emitted in  exotic particles, in units of the solar photon luminosity ($L_\odot$). Compared with those presented in Fig.~\ref{fig:exclusion}, Fig.~\ref{fig:luminosity}   provides a complementary characterization of the impact of exotic energy losses on the stellar energy budget. For   temperature dependences  ranging from constant to $T^{10}$, we find  $L_{\rm BSM} \lesssim 0.1 L_\odot$. 
The trend displayed in Fig.~\ref{fig:luminosity} reflects the interplay between the upper bounds on $\alpha$, shown in Fig.~\ref{fig:exclusion}, and the radial distribution of  energy loss. Increasing $\beta$ enhances the relative contribution of the innermost solar regions, where the temperature is the highest. 

\section{B.~Predicted neutrino fluxes\label{sec:fluxes}}

In order to investigate the impact of non-standard energy loss on the evolution of the Sun, we scan the $(\beta, \alpha)$ parameter space shown in Fig.~\ref{fig:exclusion}. We find that solar models evolved with exotic energy losses generally require slightly lower central temperatures to reproduce solar observations, see Fig.~\ref{fig:evolution}. Given the strong dependence of $^7$Be and $^8$B fluxes on the central temperature, we expect these solar models to predict lower fluxes for those components with respect to the case without energy losses ($\alpha =0$). Moreover, such models would also tend to overshoot the flux of $pp$ neutrinos due the inverse dependence of $\Phi_{pp}$. The CNO neutrino flux depends on two competing effect: the strong dependence on temperature and initial composition, particularly $Z_{\rm CNO}$, which is a free parameter in this work.

\begin{figure*}[h]
   \centering
    \includegraphics[width = \textwidth]{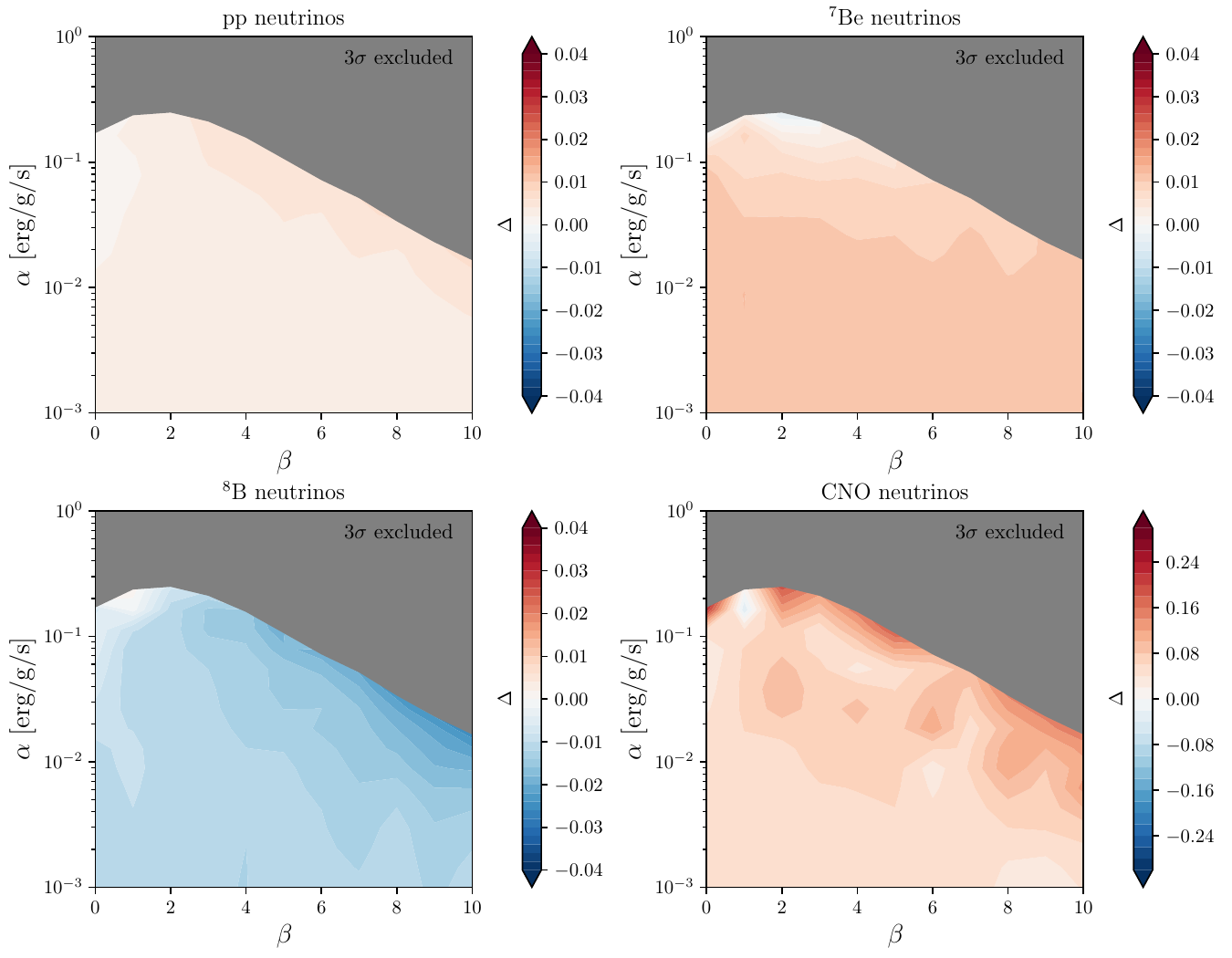}
    \caption{Relative deviation of the neutrino flux predictions for each channel ($pp$, $^7$Be, $^8$B and CNO neutrinos, from top left to  bottom right) with respect to the flux measurements in the allowed region. Relative deviations are defined as $\Delta =(\mathcal{P}_i-\mathcal{O}_i)/\mathcal{O}_i$, where $\mathcal{P}_i$ and $\mathcal{O}_i$ denote the predicted and observed fluxes, respectively. }
   \label{fig:nu-fluxes}
\end{figure*}

Figure~\ref{fig:nu-fluxes} illustrates the relative deviations of the predicted neutrino fluxes from their measured central values throughout the allowed parameter space. We define these deviations as $(\mathcal{P}_i-\mathcal{O}_i)/\mathcal{O}_i$, where $\mathcal{P}_i$ and $\mathcal{O}_i$ denote the predicted and observed fluxes, respectively. The relative deviations remain below $2\%$ for $pp$ and $^7$Be neutrinos, less than $4\%$ for $^8$B neutrinos, and $\lesssim 10\%$ for CNO neutrinos.
The different temperature and composition dependences of the neutrino components therefore provide complementary information on the solar core and play an important role in constraining non-standard energy losses. 

\section{C.~Rates of axion and hidden photon production \label{sec:comparison}}

In this Appendix, we outline the  correspondence between the  energy-loss prescription introduced in Eq.~\ref{eq:parametrisation} and the microscopic production rates often adopted in the literature. The exponent $\beta$ in Eq.~\ref{eq:parametrisation}  captures the leading temperature dependence under representative solar-core conditions, whereas the exact rates also depend on the local density, plasma frequency, screening scale, and chemical composition~\cite{Raffelt:1985nk}. Our analysis therefore constrains a general class of temperature-dependent energy-loss mechanisms that can be  translated into specific particle-production scenarios. We stress that Eq.~\ref{eq:parametrisation} does not replace a self-consistent calculation employing  the full microphysics of exotic energy losses in the solar interior, but it captures its main impact. In the following, we focus on two main scenarios considered in the main text as applications: axions and hidden photons.

\textit{Axion production through the Primakoff effect.}---Axions are hypothetical pseudoscalar particles originally introduced to solve the strong CP problem in quantum chromodynamics (QCD)~\cite{Peccei:1977hh,Weinberg:1977ma,Wilczek:1977pj}. They are also well-motivated dark-matter candidates. Through their coupling to photons, characterized by the axion-photon coupling constant ($g_{a\gamma}$), axions can be produced in stellar interiors via the Primakoff process, in which thermal photons convert into axions in the electromagnetic field of charged particles in the plasma.

The energy-loss rate per unit mass from Primakoff axion production in a nondegenerate plasma is given by~\cite{Raffelt:1985nk,Raffelt:1996wa,Vinyoles:2015aba}:
\begin{align}
    \varepsilon_{\rm axion}
    =
    \frac{g_{a\gamma}^2}{4\pi}
    \frac{T^7}{\rho}
    \mathcal{F}(\kappa^2)\, ,
    \label{eq:axion}
\end{align}
where 
$\mathcal{F}$ accounts for plasma screening effects. The screening parameter depends on the local plasma conditions through
\begin{align}
    \kappa^2 =
    \frac{\pi\alpha_{\rm em} n_B}{T^3}
    \left(Y_e+\sum_j Z_j^2Y_j\right),
\end{align}
where $\alpha_{\rm em}$ is the electromagnetic fine-structure constant, $n_B$ is the baryon number density, and $Y_j$ and $Z_j$ denote the abundance and charge of nuclear species $j$, respectively. For solar conditions, the screening function can be approximated by~\cite{Vinyoles:2015aba}
\begin{align}
    \mathcal{F}(\kappa^2)
    \simeq 1.842
    \left(\frac{\kappa^2}{12}\right)^{0.31}.
\end{align}
Its variation across the solar core is  $\lesssim 15\%$.

To estimate the normalization of the Primakoff energy-loss rate, we adopt representative solar-core conditions, with $\rho\simeq 100\,{\rm g\,cm^{-3}}$ and $\mathcal{F}\simeq 1.842$. Equation~\ref{eq:axion} then reads as
\begin{align}
    \varepsilon_{\rm axion}
    \simeq
    1.6\times10^{-3}\,
    g_{10}^2 T_7^7
    \,\,{\rm erg\,g^{-1}\,s^{-1}}\, ,
\end{align}
where $g_{10}\equiv g_{a\gamma}/(10^{-10}\,{\rm GeV}^{-1})$ and $T_7\equiv T/(10^7\,{\rm K})$.

Our generic parameterization of additional energy losses (Eq.~\ref{eq:parametrisation})  approximates the temperature dependence of Primakoff axion production for $\beta=7$, with
\begin{align}
    \alpha = 1.6\times10^{-3}g_{10}^2
    \,\,{\rm erg\,g^{-1}\,s^{-1}}\, .
\end{align}
Our $3\sigma$ bound, $\alpha<0.05\,{\rm erg\,g^{-1}\,s^{-1}}$, consequently translates into
\begin{align}
    g_{a\gamma}<5.6\times10^{-10}\,{\rm GeV}^{-1}\, .
\end{align}
This can be compared with the limit from CAST~\cite{CAST:2017uph}, $g_{a\gamma}<6.6\times10^{-11}\,{\rm GeV}^{-1}$, and the solar constraint derived in Ref.~\cite{Vinyoles:2015aba}, $g_{a\gamma}<4.1\times10^{-10}\,{\rm GeV}^{-1}$. A summary  is provided in Table~\ref{tab:limits}.
\begin{table}[h]
    \renewcommand{\arraystretch}{1.6}
     \caption{Limits at $3\sigma$ on the energy-loss normalization $\alpha$ for Primakoff axion production ($\beta=7$) and resonant emission of hidden-photon with longitudinal polarization ($\beta=1$), expressed in units of erg g$^{-1}$s$^{-1}$. We compare our results with previous solar constraints and direct searches.
     }
    \centering
        \begin{tabular}{lccc}
        \hline
        & Axion & Hidden photon & Reference \\
        \hline
        This work & 0.05 & 0.2 & -  \\
        Previous solar limits & 0.02 & 0.04 & \cite{Vinyoles:2015aba} \\
        Direct searches & 0.0007 & 0.1 & \cite{CAST:2017uph,An:2013yua} \\
        \hline
    \end{tabular}
    \label{tab:limits}
\end{table}

\textit{Resonant production of longitudinal hidden photons.}---
Hidden photons  are hypothetical massive vector bosons associated with an additional $U(1)$ gauge symmetry beyond the Standard Model~\cite{Okun:1982xi,Holdom:1985ag}. They can interact with ordinary photons through kinetic mixing, characterized by the dimensionless parameter $\chi$. Hidden photons are also viable dark-matter candidates. Unlike ordinary photons, massive hidden photons possess both transverse and longitudinal polarization modes. In particular, longitudinal hidden photons can be resonantly produced when their effective in-medium properties match those of the plasma photon modes, providing an additional energy-loss channel for stars.

The energy-loss rate per unit mass from resonant production of longitudinal hidden photons is~\cite{Vinyoles:2015aba}:
\begin{align}
    \varepsilon_{\rm HP}
    =
    \frac{\omega_p^3}{4\pi\rho}
    \frac{\chi^2m^2}{e^{\omega_p/T}-1}\, ,
\end{align}
where $\chi$ is the hidden-photon mixing parameter, $m$ is the hidden-photon mass, and $\omega_p$ is the local plasma frequency. For solar conditions, $\omega_p/T\lesssim 0.23$ (where we use  $e^{\omega_p/T}-1\simeq\omega_p/T$). The energy-loss rate then becomes
\begin{align}
    \varepsilon_{\rm HP}
    \simeq
    \frac{\omega_p^2}{4\pi\rho}
    (\chi m)^2 T\, .
\end{align}
For a representative electron fraction $Y_e=0.67$ and solar-core conditions, we obtain:
\begin{align}
    \varepsilon_{\rm HP}
    \simeq
    1.2\times10^{22}\,
    T_7\,(\chi m[{\rm eV}])^2
    \,\,{\rm erg\,g^{-1}\,s^{-1}}\, .
\end{align}

The temperature dependence of resonant longitudinal hidden-photon production is therefore approximately reproduced by Eq.~\ref{eq:parametrisation} for $\beta=1$. The corresponding normalization is
\begin{align}
    \alpha \simeq
    1.2\times10^{22}
    (\chi m[{\rm eV}])^2
    \,\,{\rm erg\,g^{-1}\,s^{-1}}\, .
\end{align}
Our $3\sigma$ bound, $\alpha<0.2\,{\rm erg\,g^{-1}\,s^{-1}}$,  translates into
\begin{align}
    \chi m<4\times10^{-12}\,{\rm eV}\, .
\end{align}
As summarized in Table~\ref{tab:limits}, this can be compared with the limit from XENON10~\cite{An:2013yua}, $\chi m<3\times 10^{-12}\,{\rm eV}$, and the solar constraint derived in Ref.~\cite{Vinyoles:2015aba}, $\chi m <1.8\times10^{-12}\,{\rm eV}$. 
As for Primakoff production, this correspondence should be considered as approximate because the microscopic rate also depends on the local plasma frequency and density, which vary throughout the stellar interior.

\end{document}